\documentclass[aps,prb,twocolumn,superscriptaddress]{revtex4-2}

\usepackage{txfonts}
\usepackage{color}
\usepackage{soul}
\usepackage{mathrsfs}
\usepackage{graphicx}

\begin{document}


\title{In-plane anisotropy of the single-$q$ and multiple-$q$ ordered phases in the antiferromagnetic metal CeRh$_2$Si$_2$ unveiled by the bulk measurements under uniaxial stress and neutron scattering}



\author{Hiraku Saito}
\email{h3110@issp.u-tokyo.ac.jp}
\affiliation{Institute for Solid State Physics, The University of Tokyo, Kashiwanoha, Chiba 277-0882, Japan}%
\author{Fusako Kon}
\author{Hiroyuki Hidaka}
\author{Hiroshi Amitsuka}
\affiliation{Department of Physics, Hokkaido University, Sapporo, Hokkaido  060-0810, Japan}%
\author{Cho Kwanghee}
\altaffiliation{Present address: Department of Physics, Chung-Ang University, Seoul 06974, Republic of Korea}
\author{Masato Hagihala}
\affiliation{Institute of Materials Structure Science (IMSS), High Energy Acceralator Research Organization (KEK), Tokai, Naka, Ibaraki 319-1106, Japan}
\author{Takashi Kamiyama}
\author{Shinichi Itoh}
\affiliation{Institute of Materials Structure Science (IMSS), High Energy Acceralator Research Organization (KEK), Tokai, Naka, Ibaraki 319-1106, Japan}
\affiliation{J-PARC Center, Tokai, Naka, Ibaraki 319-1195, Japan}
\author{Taro Nakajima}
\affiliation{Institute for Solid State Physics, The University of Tokyo, Kashiwanoha, Chiba 277-0882, Japan}%
\affiliation{RIKEN Center for Emergent Matter Science (CEMS), Wako, Saitama 351-0198, Japan}


\date{\today}

\begin{abstract}
We performed magnetization, resistivity, and neutron diffraction measurements under uniaxial stress applied along [1\={1}0] direction on the tetragonal magnet CeRh$_2$Si$_2$ with commensurate magnetic orders.
	CeRh$_2$Si$_2$ has two successive antiferromagnetic (AF) orders in zero magnetic field. %
	The high temperature phase (AF1 phase) has the magnetic modulation wave vector of $q = (\frac{1}{2}, \frac{1}{2}, 0)$, and the low temperature phase (AF2 phase) is characterized by the four $q$-vectors of $q = (\frac{1}{2}, \frac{1}{2}, 0), (\frac{1}{2}, -\frac{1}{2}, 0), (\frac{1}{2}, \frac{1}{2}, \frac{1}{2})$, and $(\frac{1}{2}, -\frac{1}{2}, \frac{1}{2})$.
	By measuring the uniaxial stress dependence of the magnetization, resistivity and the intensities of magnetic Bragg reflections, we confirmed that the AF1 phase has the single-$q$ magnetic order with two-fold rotational symmetry and the AF2 phase has the multi-$q$ magnetic order with four-fold rotational symmetry.
	In order to understand the origin of multi-$q$ order of CeRh$_2$Si$_2$, we also performed inelastic neutron scattering measurement on the single crystal samples.
	We found a magnetic excitation at the transfer energy $\hbar \omega \sim$ 8 meV.
	By applying the linear spin-wave theory, we found that the nearest and the next-nearest neighbor exchange interactions on the $ab$-plane, $J_1$ and $J_2$, are dominant in the AF2 phase. %
	However, the $J_1$-$J_2$ model cannot lift the degeneracy between the single-$q$ (AF1) and multi-$q$ (AF2) phases. %
	We suggest that it can be lifted by taking into account the biquadratic interaction derived from the perturbative expansion for the Kondo lattice Hamiltonian. [S. Hayami \textit{et al.}, Phys. Rev. B \textbf{95}, 224424 (2017).]
\end{abstract}


\maketitle

\section{Introduction}

Magnetic structures described by multiple magnetic modulation wave vectors ($q$-vectors) have attracted great interest in condensed matter physics, since the discovery of magnetic skyrmion lattices in the chiral magnet MnSi\cite{Muhlbauer,Yu2010}. 
By applying a magnetic field just below the magnetic ordering temperature, MnSi exhibits a triangular lattice of swirling spin objects, namely the triangular skyrmion lattice, on a plane perpendicular to the applied field. %
This spin texture is described by superposing three screw-type magnetic modulations and a uniform magnetization component, and thus referred to as multi-$q$ magnetic order\cite{Muhlbauer}. %
Subsequent studies discovered a variety of multi-$q$ magnetic orders\cite{Yu2011,Seki,Tokunaga,Kaneko}, some of which are accompanied by unconventional transport phenomena arising from the non-coplanar spin arrangements\cite{Neubauer,Hsu}.%

There are several possible microscopic origins to stabilize the multi-$q$ orders. %
In the early studies on the magnetic skyrmions in noncentrosymmetric magnets, a combination of ferromagnetic exchange and Dzyaloshinskii-Moriya (DM) interactions was essential to describe the helical magnetic modulations including the skyrmion lattice state\cite{Bogdanov1989,Bogdanov1994}. %
However, it was recently revealed that centrosymmetric intermetallic compounds, such as Gd$_2$PdSi$_3$ and GdRu$_2$Si$_2$, also exhibit the multi-$q$ magnetic orders, in which the DM interaction does not play a major role in determining the $q$-vector\cite{Kurumaji2017,Kurumaji2019,Khanh2020,Khanh2022,Hirschberger,Akazawa}.%
For example, GdRu$_2$Si$_2$, which is a centrosymmetric tetragonal magnet, exhibits anisotropic double-$q$, square skyrmion lattice, and spin vortex lattice states with varying magnetic field at low temperatures \cite{Khanh2022}. %
These multi-$q$ states were reproduced by calculations based on a Kondo lattice model considering couplings between conduction electrons and localized magnetic moments\cite{Ozawa,Hayami}. %

Although the multi-$q$ states are often investigated by x-ray or neutron scattering experiments, it is not straightforward to distinguish them from the multi-domain states of single-$q$ magnetic orders. %
One possible approach is to observe higher order reflections corresponding to the vector sum of the two $q$-vectors, $q_1$ and $q_2$. %
However, the higher order reflection can also be observed due to the multiple-scattering effect. %
In fact, in the previous study on the magnetic skyrmion lattice in MnSi, the azimuthal angle dependence of the reflection at $q_1+q_2$ was carefully measured to extract the intensity of the intrinsic higher order reflection\cite{Adams}. %

In the present study, we investigate magnetic orders of the intermetallic compound CeRh$_2$Si$_2$ to address the issues regarding the multi-$q$ magnetic orders. %
This system has a centrosymmetric tetragonal crystal structure\cite{Godart,Ballestracci1976,Ballestracci1978}, which is isostructural to GdRu$_2$Si$_2$. %
A previous study revealed that CeRh$_2$Si$_2$ exhibits two magnetically ordered phases in zero magnetic field\cite{Grier}, and suggested that the low temperature phase is a multi-$q$ phase\cite{Kawarazaki}.  %
We performed neutron scattering, magnetization and resistivity measurements to investigate these magnetic phases and the magnetic interactions in this system, revealing that the multi-$q$ phase of this system can also be explained by the theoretical model proposed for GdRu$_2$Si$_2$ \cite{Hayami}. %
In order to verify the multi-$q$ nature of the ground state of CeRh$_2$Si$_2$, we applied uniaxial stress to single crystal samples. %
We observed remarkable anisotropy in magnetization, resistivity and neutron diffraction intensities under uniaxial stress only in the high-temperature magnetic phase, indicating that there are irreversible changes in volume fraction of single-$q$ magnetic domains. %
By contrast, the ground state shows a robust stability against the uniaxial stress. %

\begin{figure}[t]
	\includegraphics[keepaspectratio, scale=.4]
	{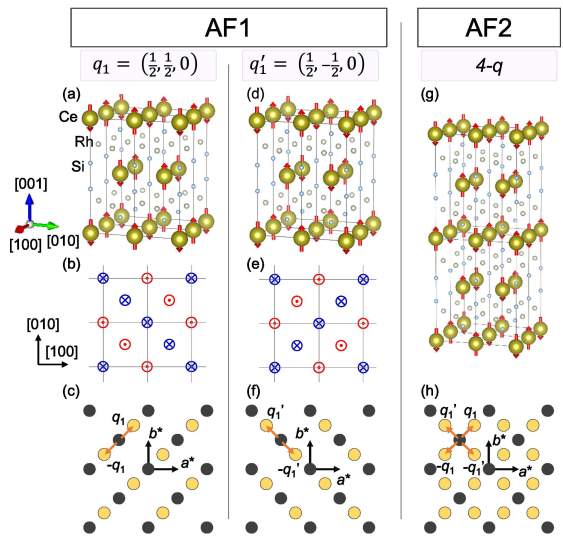}
	\caption{\label{structure}(color online) (a) Magnetic structure, (b) its projection onto $ab$ plane, and (c) the distribution of Bragg reflections in neutron experiment of $q_1$. In the same way, (a) magnetic structure, (b) its projection onto $ab$ plane, and (c) the distribution of Bragg reflections in neutron experiment of $q_1^{\prime}$. (g) Magnetic structure and (h) the distribution of Bragg reflections in neutron experiment of \textit{4-q}. Red arrows indicate magnetic moment. Black solid lines indicate unit cell in PM phase. Black(Yellow) circle indicates nuclear(magnetic) reflection. Orange arrows indicate magnetic propagation vector. }
\end{figure}

CeRh$_2$Si$_2$ crystallizes in the ThCr$_2$Si$_2$-type tetragonal structure with the space group $I4/mmm$ ($D_{4h}^{17}$, No.139)\cite{Godart,Ballestracci1976,Ballestracci1978}. %
The magnetic moments in the system are attributed to $4f$ electrons of the Ce ions\cite{Sakai}, and have strong easy-axis anisotropy along the $c$ axis\cite{Settai}. %
CeRh$_2$Si$_2$ exhibits successive antiferromagnetic transitions at $T_{\rm{N}1}$ = 36 K and $T_{\rm{N}2}$ = 25 K\cite{Grier}. %
In this paper, we refer to the high-temperature and low-temperature phases as AF1 and AF2 phases, respectively. %
According to the previous study \cite{Kawarazaki}, the AF1 phase has a single-$q$ magnetic order with a $q$-vector of $q_1 = (\frac{1}{2}, \frac{1}{2}, 0)$ as shown in Fig. \ref{structure}(a). %
In Figs. \ref{structure}(b) and  \ref{structure}(c), we show the spin arrangements of the AF1 phase projected onto the $ab$ plane and positions of magnetic Bragg peaks on the $(H,K,0)$ scattering plane in neutron diffraction experiments, respectively. %
They clearly show that the system has two-fold rotational symmetry about the $c$ axis. %
Owing to the four-fold symmetry of the crystal structure, the other $q$-vector of $q_1^{\prime} = (\frac{1}{2}, -\frac{1}{2}, 0)$ is also allowed to appear in the AF1 phase, as shown in Figs. \ref{structure}(d)-\ref{structure}(f). %
In the following, we refer to the magnetic domains corresponding to  $q_1$ and  $q_1^{\prime}$ as $q_1$-domain and $q_1^{\prime}$-domain, respectively.
In the AF2 phase, there are four $q$-vectors, specifically, $(\frac{1}{2}, \frac{1}{2}, 0), (\frac{1}{2}, -\frac{1}{2}, 0), (\frac{1}{2}, \frac{1}{2}, \frac{1}{2})$ and $(\frac{1}{2}, -\frac{1}{2}, \frac{1}{2})$\cite{Kawarazaki}, two of which are the same as the $q$-vectors in the AF1 phase. %
Figures \ref{structure}(g) and \ref{structure}(h) show the magnetic structure of the AF2 phase proposed in Ref. \onlinecite{Kawarazaki}, which is referred to as \textit{4-q}\label{key} structure, and the positions of the magnetic Bragg reflections on the $(H,K,0)$ plane, respectively. %
Despite the fact that the four-fold symmetry was broken in the AF1 phase, the proposed magnetic structure for the AF2 phase retrieves it. %
Although the remarkable change in symmetry should be observed in bulk responses such as magnetization or resistivity, they have not been investigated in detail thus far. %

\section{Experimental details}
Single crystalline samples were grown by Czochralski pulling method by using a 4-arc furnace.
They were cut into rectangular parallel-piped shape with typical size of $[1\bar{1}0] \times [110] \times [001] = 2 \times 1 \times 0.5 \, \rm{mm}^3$ except for the sample for inelastic neutron scattering (INS).
For all the measurements under uniaxial stress, $\sigma$, the direction of the uniaxial stress was fixed to be parallel to the [1\={1}0] direction.

We measured temperature dependence of the magnetization $M$ in the magnetic field of $\mu_0H$ = 0.2 T applied along the [1\={1}0] direction, which was parallel to $\sigma$.
The measurement was performed using a commercial SQUID magnetometer, Magnetic Property Measurement System (MPMS, Quantum Design inc.), with the uniaxial-stress insert used in Refs. \onlinecite{Nakajima2012,Nakajima2015}, by which we can tune the magnitude of the uniaxial stress even when the sample is at low temperatures.
Resistivity $\rho$ along [1\={1}0] direction was measured under the application of $\sigma$ using the standard four-probe method. %
The sample was mounted in the uniaxial-stress insert used in Ref. \onlinecite{Nakajima2015}, and was loaded into the Physical Property Measurement System (PPMS, Quantum Design inc.). 
Similarly to the magnetization measurements, the magnitude of the uniaxial stress was tunable at all the temperatures we measured.
Four Au-wires were attached on a (110) plane by Ag-paste with spacing each other. %
An ac current excitation with the frequency of 127 Hz was applied, and the resulting voltage was measured by a digital lock-in amplifier (LI5650, NF corp.).

The neutron diffraction measurement was performed at a triple-axis spectrometer PONTA(5G) installed in JRR-3 of Japan Atomic Energy Agency (JAEA).
The sample was mounted in a clamp-type uniaxal-stress cell so as to have the ($H, H, L$) horizontal scattering plane. %
The direction of $\sigma$ was parallel to the [1\={1}0] direction, which was normal to the scattering plane (inset of fig. \ref{ND}(a)).
The spectrometer was operated in the two-axis mode, and the horizontal collimation was $40^{\prime}-40^{\prime}-40^{\prime}$. 
An incident neutron beam with the wavelength of 2.44 \AA was obtained by a PG (002) monochromator.

INS measurement was performed at High Resolution Chopper spectrometer HRC(BL12) installed in Materials and Life science experimental facility (MLF), J-PARC\cite{Itoh}.
($H, H, L$) scattering plane was also selected for this measurement.
The measurement was performed at ambient pressure and $T$ = 7, 30, and 50 K.
Three single crystalline rods (the mass of 3.8 g in total) were coaligned, and mounted in the $^4$He closed-cycle refrigerator.
The incident pulsed neutron beam is monochromatized by a Fermi chopper with the frequency of 100 Hz.
The energy of the incident neutron $E_i$ was set to be 16.25 meV.
The energy resolution $\Delta E$ was estimated to be 1.33 meV at the elastic position.
The beam size at the sample position was 50$\times$50 mm$^2$. We employed a Soller slit collimator with a horizontal acceptance of 1.5 degree, which was installed between the Fermi chopper and the sample. 
The detector efficiency were calibrated by measurements on a vanadium standard.

The powder neutron diffraction measurement was performed at a high-resolution powder diffractometer SuperHRPD(BL08) installed in MLF, J-PARC of JAEA\cite{Torii}.
The powdered single crystal sample of a total weight of 6.84 g were installed in a vanadium cell.
The measurements were performed at $T$ = 2.5, 28, and 50 K, respectively.

\section{Results and discussions}
\subsection{Magnetization and resistivity measurements under $\sigma$}

\begin{figure}[b]
	\includegraphics[keepaspectratio, scale=.55]
	{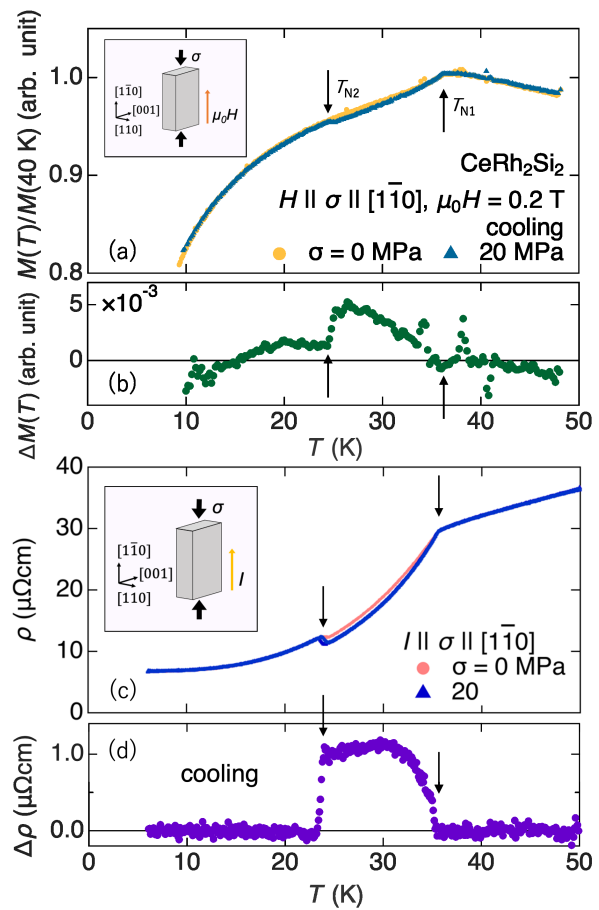}
	\caption{\label{MT} (color online) (a) Temperature dependence of magnetization $M$ at $\mu_0 H$ = 0.2 T and $\sigma$ = 0, 20 MPa. $M \,||\, H \,||\, \sigma \,||\, [1\bar{1}0]$. $M$ is normalized by $M$ at 40 K.  (b) Temperature dependence of the difference of magnetization $\Delta M$. (c) Temperature dependence of resistivity $\rho$ at $\mu_0 H$ = 0 T and $\sigma$ = 0, 20 MPa. Electric current $I$ is applied along $\sigma \,||\, \,[1\bar{1}0]$. (d) Temperature dependence of the difference of resistivity $\Delta \rho$. The measurements were performed under cooling condition. A schematic of the geometry for each measurements are shown in inset. Black arrows indicate ordering temperature.}
\end{figure}

Figure \ref{MT}(a) shows the results of the magnetization measurements in an external magnetic field of 0.2 T applied parallel to the $[1\bar{1}0]$ direction. %
We first measured the $M$-$T$ curve with decreasing temperature at ambient pressure. %
$M$ shows a cusp anomaly at $T_{\rm{N}1}$ = 36 K. %
The system underwent the magnetic phase transition at $T_{\rm{N}2}$, which did not show clear anomaly in the present $M$-$T$ measurement at ambient pressure. %
We then applied $\sigma$ = 20 MPa parallel to the $[1\bar{1}0]$ direction as shown in the inset of Fig. \ref{MT}(a) at 48 K, and measured $M$-$T$ curve on cooling. %
We found that the anomaly at $T_{\rm{N}1}$ was not affected by the application of $\sigma$. %
However, the magnetization in the AF1 phase was suppressed by $\sigma$. %
We also found that the data at $\sigma$ = 20 MPa showed a step-like change at $T_{\rm{N}2}$, below which the data nearly coincides with that measured at ambient pressure. %
These observations show that the application of $\sigma$ leads to the reduction of $M$ only in the AF1 phase. %
This can also be seen in the temperature dependence of $\Delta M$, which is the difference between magnetizations measured with $\sigma$ = 0 and 20 MPa, shown in Fig. \ref{MT}(b).
Note that the magnetization data are normalized to the data measured at 40 K in each cooling run, assuming that the magnetization does not depend on $\sigma$ in the paramagnetic phase. %
This assumption is justified by measuring $\sigma$ dependence of $M$ at fixed temperatures, as we mention shortly. %
We also note that the remarkable downturns of the magnetization at low temperatures are the temperature dependent background signals of the uniaxial-stress cell. %

We also performed resistivity measurements under $\sigma$ in zero magnetic field. %
Similarly to the magnetization measurements, the uniaxial stress was applied along the $[1\bar{1}0]$ direction. The direction of the electric current was selected to be parallel to $\sigma$ as depicted in the inset of Fig. \ref{MT}(c). %
At ambient pressure, $\rho$ shows a kink anomaly and a step-like change at $T_{\rm{N}1}$ and $T_{\rm{N}2}$, respectively. %
By the application of $\sigma$ = 20 MPa (blue triangle), $\rho$(20 MPa) becomes smaller than $\rho$(0 Pa) (pink circle) only in the AF1 phase. %
The difference between $\rho$ measured at 0 and 20 MPa, $\Delta \rho$, shows dramatic increase below $T_{\rm{N}1}$ and drops to zero at $T_{\rm{N}2}$ (Fig. \ref{MT}(d)). %
It indicates that the electric property of CeRh$_2$Si$_2$ is also sensitive to $\sigma$ along [1\={1}0] in the AF1 phase, and insensitive in the AF2 phase.

\begin{figure}[t]
	\includegraphics[keepaspectratio, scale=.65]
	{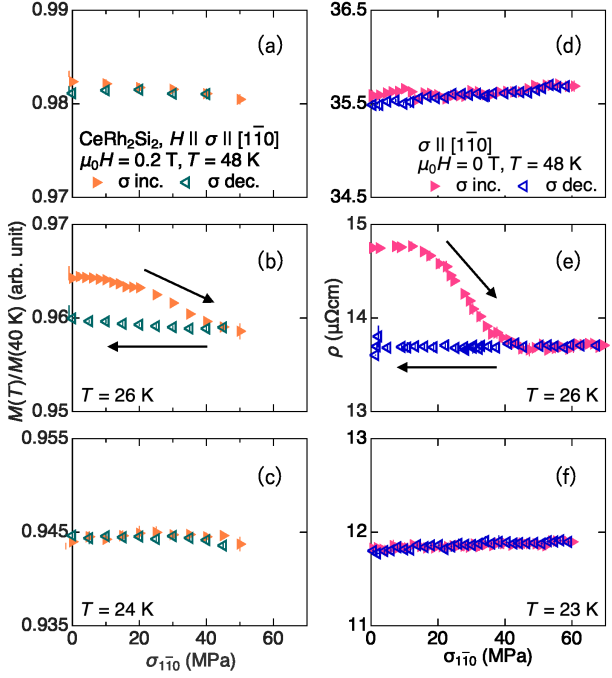}
	\caption{\label{sigdep} (color online) (a)-(c) Uniaxial stress dependence of $M$ at $\mu _0 H$ = 0.2 T and $T$ = 48, 26, 24 K, respectively. $M \,||\, H\,||\,\sigma\,||\,[1\bar{1}0]$. $M$ is normalized by $M$ at 40 K. (d)-(f) Uniaxial stress dependence of $\rho$ at $T$ = 48, 26, and 23 K, respectively. $I \,||\, \sigma\,||\,[1\bar{1}0]$. Closed(open) symbols indicates the data measured under $\sigma$ increase(decrease). Black arrows are guides to the eye.}
\end{figure}

We further investigate the $\sigma$ dependence of $M$ and $\rho$ by measuring their $\sigma$ dependence at fixed temperatures, as shown in Figs. \ref{sigdep}(a)-\ref{sigdep}(f). %
As mentioned in the introduction, we expect the formation of the multi-domain state in the AF1 phase. %
The fractions of the $q_1$ and $q_1^{\prime}$ domains are supposed to be sensitive to an anisotropic perturbation which breaks the four-fold rotational symmetry of the crystal. %
The change in fraction of the single-$q$ magnetic domains should be accompanied by dissipative motions of magnetic domain walls. %
Indeed, we observed large irreversible changes of $M$ and $\rho$ only in the AF1 phase, while there are no significant $\sigma$ dependence in the paramagnetic and AF2 phases. %
These results confirmed that the AF1 phase has the multi-domain state of the single-$q$ magnetic domains, and revealed that the AF1 phase has magnetic and electronic anisotropy reflecting its two-fold rotational symmetry, which was not directly observed in the previous study. %
This also highlights that the AF2 phase exhibits isotropic natures in terms of both electronic and magnetic properties, which supports the multi-$q$ magnetic order with four-fold rotational symmetry. %

\subsection{Neutron diffraction measurements}
\begin{figure}[ht]
	\includegraphics[keepaspectratio, scale=.6]
	{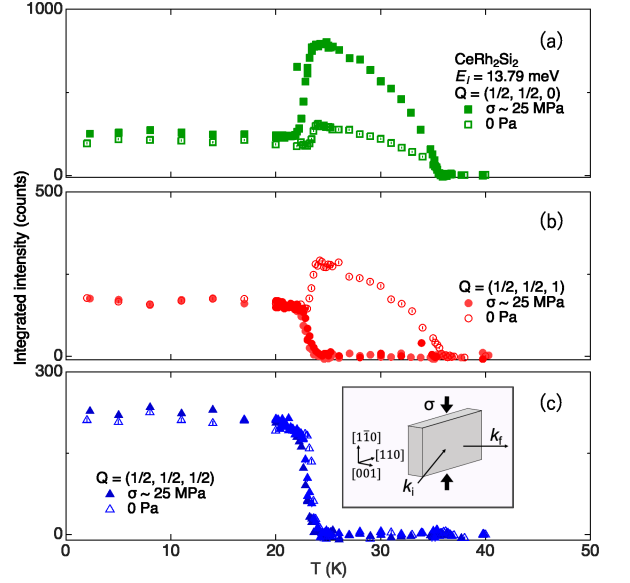}
	\caption{\label{ND} (color online) Temperature dependence of integrated intensity at $\sigma$ = 0 and $\sim$25 MPa at $Q=$ (a) $(\frac{1}{2}, \frac{1}{2}, 0)$, (b) $(\frac{1}{2}, \frac{1}{2}, 1)$, and (c) $(\frac{1}{2}, \frac{1}{2}, \frac{1}{2})$, respectively. Open(Closed) symbols are the data at $\sigma$ = 0 Pa($\sim$25 MPa). The experimental geometry is shown in inset of (c). $k_i$ and $k_f$ indicates the incident and reflection neutron, respectively.}
\end{figure}

\begin{figure}[ht]
	\includegraphics[keepaspectratio, scale=.55]
	{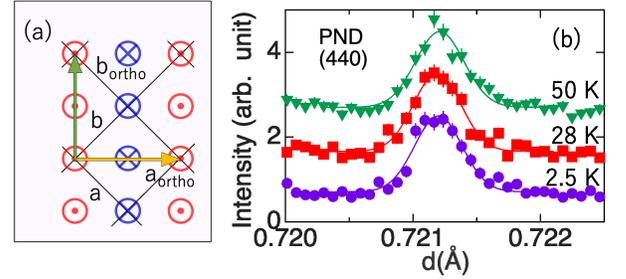}
	\caption{\label{PND} (color online) (a) A schematic projection onto $ab$-plane of the geometric relation between tetragonal and orthorhombic lattices in AF1 phase. The blue and red arrows pointing normal to plane indicates magnetic moments. The solid lines indicate tetragonal unit cell. Yellow and green arrow indicates translation vector corresponding to orthorhombic lattice. The constants $a$ and $b$ ($a_{ortho}$ and $b_{ortho}$) indicates lattice constants of tetragonal(orthorhombic) unit cell. (b) The peak profiles of (440) reflection obtained by powder neutron diffraction measurement at $T$ = 50, 28, and 2.5 K.}
\end{figure}

To directly observe the volume fractions of the single-$q$ magnetic domains, we performed single-crystal neutron diffraction measurements at ambient pressure and under $\sigma$ of $\sim$25 MPa applied parallel to the $[1\bar{1}0]$. %
On the $(H,H,L)$ scattering plane, we focus on two magnetic Bragg reflections appearing at $Q = (\frac{1}{2}, \frac{1}{2}, 0)$ and $(\frac{1}{2}, \frac{1}{2}, 1)$. %
By taking into account the reflection condition for the body-centered lattice, these reflections can be indexed as $\tau+q_1$ and $\tau+q_1^{\prime}$, respectively, where $\tau$ is a vector to a reciprocal lattice point satisfying $H+K+L=2n$ ($n$ is an integer). %
Thus, the integrated intensities of these reflections are proportional to the volume fractions of $q_1$ and $q_1^{\prime}$ domains in the AF1 phase. %
The temperature dependence of the integrated intensity of $Q = (\frac{1}{2}, \frac{1}{2}, 0)$, and $(\frac{1}{2}, \frac{1}{2}, 1)$ are shown in Figs. \ref{ND}(a) and \ref{ND}(b), respectively. %
The open and closed symbols indicate the data measured at $\sigma$ = 0 Pa and $\sim$25 MPa, respectively. %
At ambient pressure, both the reflections start to grow at $T_{\rm{N}1}$ with decreasing temperature. %
However, under $\sigma$ of $\sim$ 25 MPa, the intensity at $(\frac{1}{2}, \frac{1}{2}, 0)$ significantly increased in the AF1 phase, and instead, the reflection at  $(\frac{1}{2}, \frac{1}{2}, 1)$ completely disappeared. %
This unambiguously shows that the application of $\sigma = \, \sim$25 MPa along the $[1\bar{1}0]$ direction leads to the single-domain state with the $q_1$ domain in the AF1 phase. %

Similarly to the magnetization and resistivity measurements, the intensities at $(\frac{1}{2}, \frac{1}{2}, 0)$ and $(\frac{1}{2}, \frac{1}{2}, 1)$ measured under $\sigma$ coincide with the data measured at ambient pressure below $T_{\rm{N}2}$. %
We also measured the magnetic Bragg reflection at  $(\frac{1}{2}, \frac{1}{2}, \frac{1}{2})$, which is characteristic of the AF2 phase. %
As shown in Fig.  \ref{ND}(c), this reflection was hardly affected by the application of $\sigma$. %
This is another evidence that the AF2 phase has the multi-$q$ structure which is not affected by the relatively weak uniaxial stress applied along the in-plane direction. %

As for the AF1 phase, the fact that the magnetic domains are controlled by the application of the uniaxial stress means that the magnetic domains with the $q$-vector perpendicular to $\sigma$ gains elastic energy, implying that there could be tiny lattice distortions associated with the spin arrangements in the AF1 phase. %
Specifically, the present results suggest that the symmetry of the crystal structure in the AF1 phase is orthorhombic, and that the orthorhombic $a$ axis ($a_{ortho}$), which is defined to be parallel to the $q$-vector, is supposed to be slightly longer than the orthorhombic $b$ axis ($b_{ortho}$), as illustrated in Fig. \ref{PND}(a).
This would result in splitting (or broadening) of nuclear Bragg peak with relatively large $H$ and $K$ indices. We thus performed high-resolution neutron powder diffraction measurements at SuperHRPD (BL08) in MLF of J-PARC. %
However, we did not observed any peak splitting nor broadening. For instance, we show the temperature dependence of the 440 nuclear Bragg reflection, which is indexed by the original tetragonal cell, in Fig. \ref{PND}(b).  
These results suggest that the possible lattice distortion could be smaller than the accuracy of the measurement, specifically $\delta d/d \sim 9.0 \times 10^{-5}$, if exists. %

\subsection{Inelastic neutron scattering}

\begin{figure}[b]
	\includegraphics[keepaspectratio, scale=.45]
	{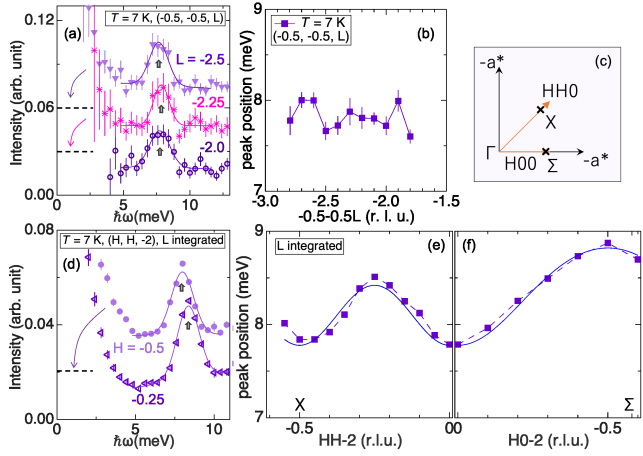}
	\caption{\label{INS} (color online) (a) Inelastic spectrum at $T$ = 7 K at Q = (-0.5, -0.5, $L$) for $L$ = -2.0, -2.25, and -2.5. The solid lines are fitting curve. The peak positions are indicated by arrows. Data are shifted vertically. The level is shown by dashed line. (b) $L$ dependence of peak position along (-0.5, -0.5, $L$). (c) A schematic of the data trajectory shown in Fig. 6. $\Gamma$, $\textrm{X}$, and $\Sigma$ indicate critical points of Brillouin zone. (d) Inelastic spectrum at $T$ = 7 K at Q = ($H$, $H$, -2) for $H$ = -0.5 and 0.25. Note that the data are integrated along $L$. (e), (f) $H$ dependence of peak position at $T$ = 7 K along ($H$, $H$, -2) and ($H$, 0, -2), respectively. The solid line are fitted curve based on the linear spin-wave model.}
\end{figure}
\begin{figure}[b]
	\includegraphics[keepaspectratio, scale=.35]
	{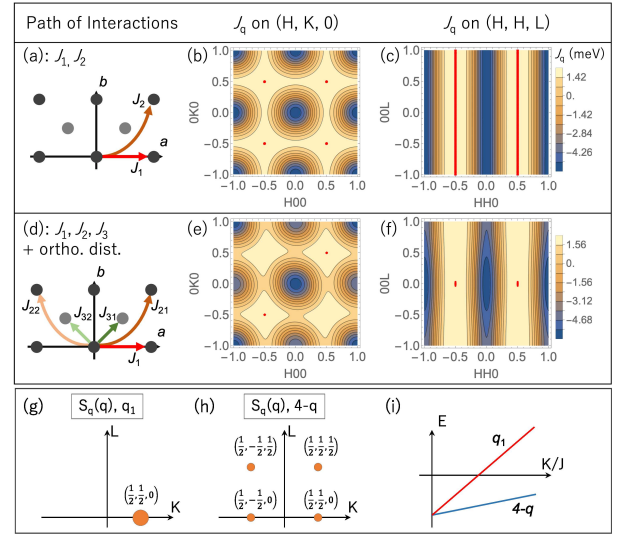}
	\caption{\label{Jq} (color online) (a) The path of interactions, $J_1$ and $J_2$. Black circle indicates Ce ions in the plane. Gray circle indicates Ce ions 1/2z above plane. (b), (c) The contour maps of $J_q$ for (a) in ($H$, $K$, 0) and ($H$, $H$, $L$) plane, respectively. (d) The path of interactions, $J_1$, $J_{2}$, and $J_{3}$. By assuming the orthorhombic distortion shown in Fig. \ref{Jq}(a), $J_2(J_3)$ would split into $J_{21}(J_{31})$ and $J_{22}(J_{32})$. We assume that $J_3$ is antiferromagnetic interaction\cite{comment1} with the magnitude 4 times smaller than $J_2$. The degree of splitting is assumed to be $J_{22}/J_{21} = J_{32}/J_{31}$ = 3/4. (e), (f) The contour maps of $J_q$ for (d) in $(H, K, 0)$ and $(H, H, L)$ plane, respectively. (g), (h) The schematics of $S_q$ in $(\frac{1}{2}, K, L)$ plane for $q_1$ and \textit{4-q}, respectively. The size of circle corresponds to the strength of $S_q$. (i) A schematic $K/J$ dependence of the energy $E$ for bilinear-biquadratic model on CeRh$_2$Si$_2$.}
\end{figure}

We measured magnetic excitation spectra at $T$ = 7 K in the AF2 phase at HRC. %
Energy spectra along (-0.5, -0.5, $L$) at $T$ = 7 K for several $L$ are shown in Fig. \ref{INS}(a). %
The data are shifted vertically with each other. %
We found a magnetic excitation at around $\hbar \omega \sim$ 8 meV. %
That is consistent with a previous neutron scattering measurement on a powder sample\cite{Willers}.
We extracted the excitation energy as function of $L$ (Fig. \ref{INS}(b)) and found no systematic $L$ dependence within the accuracy of the present experiment. %
This indicates that the magnetic interactions along the $c$ axis are significantly weak as compared to those in the $ab$ plane. %
In the following, we thus integrated the data along 00L  (Figs. \ref{INS}(d)-\ref{INS}(f)) to get better statistics. %
The energy spectra at ($H, H$, -2) where $H$ = -0.5 and -0.25 are shown in Fig. \ref{INS}(d).
The integration range of $L$ is from -1.9 to -2.7.
The data apparently show that the peak position depends on $H$. %
We extracted the excitation energies along the ($H, H,$ -2) and ($H$, 0, -2) lines shown in Fig. \ref{INS}(c), and obtained the dispersion relations of the magnetic excitations as shown in Figs. \ref{INS}(e) and \ref{INS}(f), respectively. %
These dispersion relations can be well reproduced by calculations based on the linear spin-wave theory\cite{Squires} including the nearest and next-nearest exchange interactions, $J_1$ and $J_2$, and the uniaxial anisotropy $D$. %
From the fitting analysis, these parameters are estimated to be $J_1 = -0.98$ meV, $J_2 = -0.40$ meV, and $D = 18.0$ meV, respectively (solid line in Figs. \ref{INS}(e) and \ref{INS}(f)).
It indicates that $J_1$ and $J_2$ are dominant in CeRh$_2$Si$_2$.
We also performed neutron inelastic scattering measurements in the AF1 phase, and found that the dispersion relations are qualitatively the same as those in the AF2 phase (see supplemental).

We calculated Fourier transform of the exchange interactions $J_q = \sum_{j} J_{i,j}e^{-iq \cdot (r_i -r_j)}$. %
The results are shown as a contour map in ($H, K$, 0) (Figs. \ref{Jq}(b) and \ref{Jq}(e)), and ($H, H, L$) planes (Figs. \ref{Jq}(c) and \ref{Jq}(f)). %
The maximum of $J_q$ are indicated by red points and lines. 
We start with a minimal model including only $J_1$ and $J_2$.
The path of interaction is shown in \ref{Jq}(a).
The parameters are fixed at the estimated value obtained from INS.
On the $(H, K, 0)$ plane, there are four maxima corresponding to $q_1$, $q_1^{\prime}$, $-q_1$, and $-q_1^{\prime}$. %
Reflecting the absence of the out-of-plane magnetic interactions, the energies at these four points are independent of $L$ as shown in Fig. \ref{Jq}(c). %
In this situation, the spin arrangements of the AF1 and AF2 phases give the same energy, because the in-plane spin arrangements in these phases are identical to each other. %
We thus need to introduce an additional parameter to lift the degeneracy and to realize the multi-$q$ state. %

Here, we employ the theoretical model used in the previous study on GdRu$_2$Si$_2$\cite{Hayami}. %
They applied the bilinear-biquadratic model, which was derived from a Kondo lattice model, described by $\mathscr{H} = \Sigma_q [-J\langle S_q\rangle  \cdot \langle S_{-q}\rangle + K(\langle S_q\rangle  \cdot \langle S_{-q}\rangle)^2]$.
$\langle S_q\rangle$ is Fourier transformed spin spin component with the wave vector $q$. %
The first term corresponds to $J_q$ calculated in Figs. \ref{Jq}(a)-(c). %
The second term arises from the higher-order coupling between the conduction electrons and localized magnetic moments. %
By introducing the finite $K$ parameter, the degeneracy between the single-$q$ state with $q_1$ and the \textit{4-q} state is immediately lifted. %
Specifically, in the single-$q$ state, the Fourier component is finite only when $q=q_1$ (Fig. \ref{Jq}(g)). %
However, in the \textit{4-q} state, the Fourier components are distributed to the four points, namely $(\frac{1}{2}, \frac{1}{2}, 0),  (\frac{1}{2}, -\frac{1}{2}, 0),  (\frac{1}{2}, \frac{1}{2}, \frac{1}{2})$ and  $(\frac{1}{2}, -\frac{1}{2}, \frac{1}{2})$ (Fig. \ref{Jq}(h)), and the amplitudes of the components are smaller than that of the single-$q$ state. %
When the sign of $K$ is positive, the second term, which is called biquadratic term, costs the energy in both cases. But the single-$q$ state is more destablized than the \textit{4-q} state (Fig. \ref{Jq}(i)). %
We thus conclude that the formation of the multi-$q$ state in CeRh$_2$Si$_2$ can also be qualitatively understood by the bilinear-biquadratic model applied for GdRu$_2$Si$_2$. %

We note that the energy cost due to the $K$ term would be further suppressed by introducing a multi-$q$ order with incommensurate magnetic modulations along the $c^*$ direction.
However, we did not observed any indication of it by neutron diffraction.

As for the AF1 phase, we experimentally confirmed the single-$q$ magnetic order, which could be associated with a tiny lattice distortion. 
One of the possible microscopic model to explain the spin-lattice coupling is the 3rd neighbor exchange interaction, $J_3$, with the orthorhombic distortion as shown in Fig. \ref{Jq}(e). 
The contour maps of $J_q$ are modified by the anisotropic 3rd neighbor interactions as shown in Figs. \ref{Jq}(e) and (f); the degeneracy between $q_1$ and $q_1^{\prime}$ is lifted and the maxima appear on the $L$ = 0 plane.
Another possibility is in-plane anisotropy of the 4$f$ electron distribution of the Ce ion. 
The anisotropy may be induced in the $\Gamma^{(1)}_7$ ground state of CeRh$_2$Si$_2$ \cite{Patil,Abe,Amorese,Willers} by applying the uniaxial stress, which could favor either the $q_1$ or $q_1^{\prime}$ domain. 
To confirm these scenarios, further investigations such as neutron inelastic scattering measurements in the single-domain AF1 phase or detailed magnetoelastic/magnetostriction measurements would be necessary.  

Finally, we would mention that the driving force of the thermally-induced phase transition from the multi-$q$ AF2 phase to the single-$q$ AF1 phase remains to be understood. 
A previous theoretical study by K. Barros and Y. Kato\cite{Barros2013} demonstrated similar multi-$q$-to-single-$q$ transitions using a Kondo lattice model, although the origins of the transitions were not elucidated. 
To quantitatively compare the stabilities of the multi-$q$ and single-$q$ phases at finite temperatures, further theoretical studies would be necessary.

\section{summary}
To summarize, we have performed magnetization, resistivity, and neutron diffraction measurement under the application of the uniaxial stress, and inelastic neutron scattering measurement on single crystalline CeRh$_2$Si$_2$.
We confirmed that the AF1 phase has the single-$q$ magnetic structure with the 2-fold rotational symmetry, and that the system is in the multi-domain state at ambient pressure. 
The results imply that there is a tiny lattice distortion, though it cannot be detected in the accuracy of $\delta d/d \sim 9.0 \times 10^{-5}$.
By contrast, the AF2 phase is stubborn to in-plane uniaxial stress and recover the 4-fold rotational symmetry reflecting the multi-$q$ state.
We found that CeRh$_2$Si$_2$ shows dispersive magnetic excitation at $\hbar \omega \sim$ 8 meV.
A fitting based on the linear spin-wave model shows $J_1$ and $J_2$ are dominant in CeRh$_2$Si$_2$.
We revealed that $K$ term of the bilinear-biquadratic model lifts the degeneracy of $q_1$ and \textit{4-q} in CeRh$_2$Si$_2$.
It implies the applicability of the model to the commensurate multi-$q$ magnetic orders.
To understand the mechanism of the thermally-induced multi-$q$-to-single-$q$ magnetic phase transition in finite temperature, it would be necessary to evaluate quantitatively the biquadratic interaction and effects of thermal fluctuations.
Finally, we also note that an application of a relatively weak uniaxial (anisotropic) stress could be a useful tool to distinguish multi-$q$ orders from (trivial) single-$q$ orders.

\begin{acknowledgments}
The authors are grateful to S. Hayami for fruitful discussion.
A part of the magnetization and resistivity measurements were performed by using PPMS and MPMS at the user laboratory of the Comprehensive Research Organization for Science and Society.
The neutron scattering experiment at JRR-3 was carried out along the proposal No. 21513 and partly supported by ISSP of the University of Tokyo.
The neutron experiment performed at SuperHRPD, MLF, J-PARC was performed under a user program (Proposal No. 2019B0083).
The neutron experiment performed at HRC were approved by the Neutron Scattering Program Advisory Committee of IMSS, KEK (No. 2018S01 and No. 2019S01).

\end{acknowledgments}

\bibliography{CeRh2Si2_uniax_aps}

\end{document}